\documentclass{aastex61}

\received{August 2, 2017}
\revised{February 19, 2018}
\accepted{March 12, 2018}
\submitjournal{AJ}

\shorttitle{SFR distribution in NGC~1232}
\shortauthors{ Souza et al.}

\begin{document}

\title{Star formation rate distribution in the galaxy NGC~1232}

\correspondingauthor{Alexandre Ara\'ujo de Souza}
\email{ale.plie@hotmail.com}

\author{Alexandre Ara\'ujo de Souza}
\affil{Universidade Cruzeiro do Sul \\
Rua Galv\~ao Bueno, 868 \\
S\~ao Paulo, SP. CEP 01506-000. BR}
\affiliation{CRAAM, Mackenzie Presbyterian University \\
Rua da Consolacao, 896\\
Sao Paulo, Brazil}

\author{Lucimara P. Martins}
\affiliation{Universidade Cruzeiro do Sul \\
Rua Galv\~ao Bueno, 868 \\
S\~ao Paulo, SP. CEP 01506-000. BR}

\author{Alberto Rodr\'iguez-Ardila}
\affiliation{Laborat\'orio Nacional de Astrof\'isica\\
R. dos Estados Unidos, 154 - Nações\\
Itajub\'a - MG, 37530-000. BR}

\author{Luciano Fraga}
\affiliation{Laborat\'orio Nacional de Astrof\'isica\\
R. dos Estados Unidos, 154 - Nações\\
Itajub\'a - MG, 37530-000. BR}



\begin{abstract}

NGC~1232 is a face-on spiral galaxy and a great laboratory for the study of star-formation due 
to its proximity. We obtained high spatial resolution H$\alpha$ images of this galaxy, with 
adaptive optics, using the SAM instrument at the SOAR telescope, and used these images to 
study its H\,{\sc ii} regions. These observations allowed us to 
produce the most complete H\,{\sc ii} region catalog for it to date, with a total of 976 sources. 
This doubles the number of H\,{\sc ii} regions previously found for this object. 
We used these data to construct the H\,{\sc ii} luminosity function, and obtained a power-law index 
lower than the typical values found for Sc galaxies. This shallower slope is related to the 
presence of a significant number of high-luminosity H\,{\sc ii} regions (logL$>$39 dex). 
We also constructed the size distribution function, verifying that, as for most galaxies, 
NGC~1232 follows an exponential law. We also used the H$\alpha$ luminosity to calculate the star 
formation rate. An extremely interesting fact about this galaxy is that 
X-ray diffuse observations suggest that NGC~1232 recently suffered a collision with a 
dwarf galaxy. We found an absence of star formation around the region where the X-ray emission 
is more intense, which we interpret as a star formation quenching due to the collision. 
Along with that, we found an excess of star-forming regions in the northeast part of the galaxy, 
where the X-ray emission is less intense.

\end{abstract}




\section{Introduction} \label{sec:intro}


The distribution of H\,{\sc ii} regions is an excellent tracer of the star formation in spiral galaxies
\citep[e.g.][]{kennicutt+92}. Catalogs of these regions provide an observational base that can be 
used to study global star formation across galactic disks. From observational properties of these
regions  (e.g., recombination lines or dust emission) quantities like the star formation rate (SFR) can
be obtained.

The SFR is an important factor in the chemical evolution of a galaxy. Its value 
gives the total amount of gas converted to stars over a given time interval, which may depend on 
several environmental properties. It is long known that a large fraction of the star formation in the local 
universe occurs in gas-rich disk galaxies \citep{kennicutt+89}. Optical imaging clearly reveals that 
spiral arms in disk galaxies have a high concentration of young stars, implying that SFR must be higher 
in the arms regions than anywhere else in the galaxy. However, it is still a matter of debate if the 
higher SFR in the arms is just an effect of the larger gas densities  \citep[e.g.][]{elmegreen+86,elmegreen95,foyle+10}, 
or if the mass excess in the arms could directly act to trigger the star 
formation 
\citep[e.g.][]{roberts+75,gittins+04,seigar+02,gittings+04}. 

Grand-design and multi-arm spiral galaxies are ideal laboratories to study these effects. 
However, there 
are only a few of these galaxies close enough that individual clusters can be analyzed. A 
high-resolution 
study of these systems can help to shed some light on the nature of the processes converting gas into stars. 

NGC~1232 is technically a grand-design spiral galaxy, practically face-on
\citep[inclination$=$29$^0$,][]{devaucouleurs+91,corwin+94}. Morphologically, it
is classified as a SAB(rc)c. It has well-defined spiral arms,
despite being a bit uncommon as they do not wind smoothly as expected for a galaxy of 
this type. Its arms appear to be ``bent" in areas rather than gently winding structures seen in
undisturbed spiral galaxies, which is probably related to the gravitational distortion 
caused by its satellite \citep{arp82}.  It is considered 
a prototype of multi-arm regular spirals, having hints of a bar in the nuclear region, a small bulge and 
long arms that disperse to the external regions, producing a number of thin arms. Radio studies suggest that 
it has a large neutral gas envelope that extends much beyond the optical limits of the galaxy 
\citep{vanzee+99}. It also has a satellite galaxy, NGC~1232A, with which it is believed it has interacted 
in complex ways, 
but 
due to the large difference in their redshifts and lack of any other sign of
physical association, it is likely not currently interacting with \citep{arp82}.  
The estimated distance between NGC~1232 and its satellite is 2.4~Mpc \citep{vanzee+99},
which means that they are not currently physically associated. 
Both galaxies may be associated with the Eridanus group of galaxies, but 
at a projected distance of 2.2~Mpc from the center of the low-mass cluster, neither 
is likely to be bound to the cluster \citep{Willmer+89}.

NGC~1232 covers about 6'.7 $\times$ 7'.8 on the sky at a distance of 19.8 Mpc\footnote{Distance
obtained from NED (NASA/IPAC Extragalactic Database)}. The nuclear region seems dominated by an older 
population, as evidenced from the spectra in the optical \citep[6df galaxy survey;][]{heath+09} and 
near-infrared \citep{martins+13}, while the spiral arms are populated by 
numerous regions of star formation.  

In a recent study, \cite{garmire+13} provides evidence that NGC~1232 suffered a collision with a 
dwarf galaxy using X-ray images from NASA Chandra Observatory. This result is based on the detection of an 
unusual pattern of diffuse emission, which is related to a shocked gas region with a temperature of
5.8~MK, covering an impact area of 7.25 kpc in diameter. The image of the collision (Figure 1 of Garmire 2013) 
reveals a cloud with a cometary appearance sweeping across the galaxy and possibly colliding with the disk. 
The center of the collision is about 4.3~kpc to the west of the nucleus of NGC~1232. According to the authors 
the shock wave of this collision may have triggered some star formation, producing bright and massive stars. 

Motivated by the observed diffuse X-ray emission by \cite{garmire+13},
we obtained optical high angular resolution imaging of NGC~1232 in
g', r', and H$\alpha$ to study its star-formation activity, searching for the connection to the spiral
arms and looking for additional evidence of this putative collision. The paper is divided as follows: in
\S 2, we present the observations and data reduction; 
in \S 3, we present the H\,{\sc ii} regions detection; 
in \S 4 we present the catalog properties; in \S 5 we obtain the SFR distribution and discuss the results; 
and in \S 6 we present our 
conclusions.

\section{Observations and Data reduction}

\subsection{ Observations}

NGC~1232 was observed on the night of 2014 January 23 (Program SO2013B-021), with the filters g', r', and H$\alpha$. 
The observations were carried out with the SOAR telescope using the SOAR Adaptive Module (SAM), 
an instrument with an installed laser-assisted adaptive optics (AO) system \citep{fraga+13}. By selectively 
compensating for low-altitude turbulence, AO improves resolution at visible 
wavelengths. The instrument contains a 4k $\times$ 4k pixels CCD sensor that spans a square field of 3$\arcmin$.
The natural seeing in the V band during the observations was 0.8$\arcsec$. With AO the V band seeing 
improved to 0.55$\arcsec$. In H$\alpha$ the seeing was 0.5$\arcsec$ and in the g' band, 0.66$\arcsec$. Three individual 
on-source integrations were carried out for each filter, of 240~s each for g' and r' and 
600~s for H$\alpha$. The standard star Hiltner~600 was also observed in the three filters for flux calibration
purposes. 

\subsection{Reduction and Calibration}

The images were reduced in a standard way using the Image Reduction and Analysis Facility 
(IRAF\footnote{IRAF is distributed by the National Optical Astronomy Observatory,
which is operated by the Association of Universities for Research in Astronomy, Inc., under
cooperative agreement with the National Science foundation.};\cite{valdes98,valdes+98}). The reduction 
process includes bias subtraction, flat-field corrections, and cosmic-ray cleaning.
Correction for Galactic extinction was applied using the dust maps from \citet{schlegel+98}
and the extinction law from \citet{cardelli+89}. 
SAM has a substantial optical distortion, typical of two off-axis parabolic mirrors \citep{fraga+13}. 
The distortion can reach up to 42 pixels (1.93" in the sky) in the image. The distortion is well described by 
quadratic terms in $x$ and $y$ (in pixels, described in the instrument manual), 
and we applied these corrections to our images.

\begin{figure*}[h]
\centering
\includegraphics[scale=0.9]{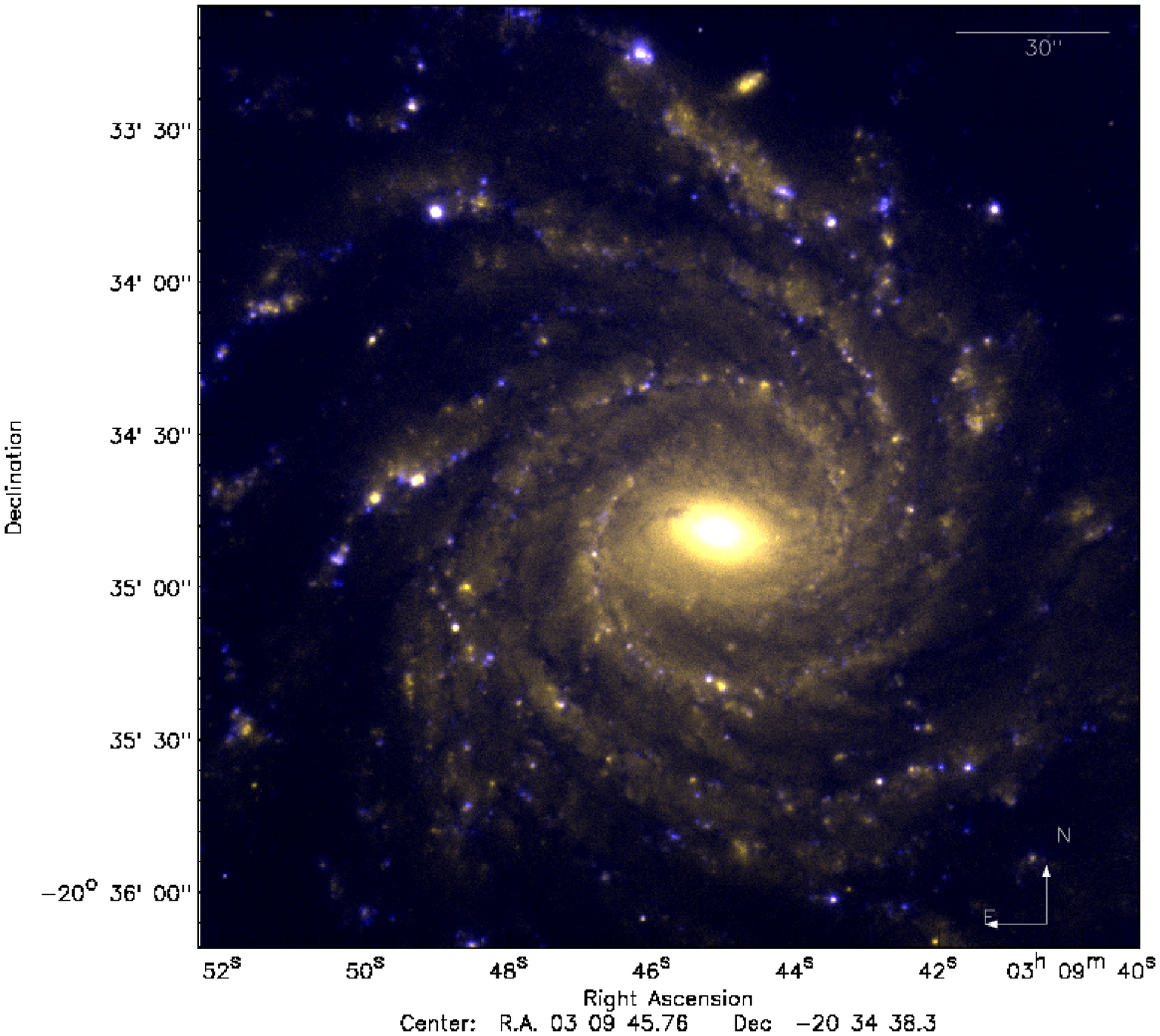} 
\caption{NGC~1232 combined
flux-calibrated} image of the filters g'(green), r'(red),
and H$\alpha$(blue), obtained with the instrument SAM at SOAR.
\label{ngc1232_filters}
\end{figure*}

Flux calibration was done using the g', r' and H$\alpha$ images of the standard star. Using these
images, the transmission curves of each filter\footnote{Obtained at http://www.ctio.noao.edu/soar/content/filters-available- Soar} 
and its flux-calibrated spectrum obtained from \citet{Hamuy+92,Hamuy+94},
we calculated the conversion factor from photon counts to flux units (erg~cm$^{-2}$~s$^{-1}$). 
Figure~\ref{ngc1232_filters} shows the resulting image obtained after combining the final
exposures in each filter. 

\section{ H$\alpha$ photometry and H\,{\sc ii} region identification}

The continuum subtraction was done using the r' image as described in \citet{schmitt+06}. 
Since this is a broadband filter and encompasses the H$\alpha$ emission-line region, 
it is contaminated by that emission. This means that besides the scaling for the 
H$\alpha$ filter width, it has to be corrected for the contribution of the H$\alpha$ flux to 
the total observed flux. The subtraction then has to be recursive. 
This is done by first subtracting the scaled continuum image from the line image; then, the
resulting line image was subtracted from the continuum image to remove the emission-line contribution 
to this image, and the corrected continuum image was used to subtract the
continuum emission from the original line image. This process was
repeated a few times, until the H$\alpha$ and continuum fluxes in regions
affected by contamination did not change, indicating that the process had converged. In total, 
we needed six iterations to reach convergence. 

The detection of the H\,{\sc ii} regions in the NGC~1232 H$\alpha$ image was performed with the aid of 
SExtractor \citep{bertin+96}. 
This software is used to build a catalog of objects from an astronomical image.
It first determines the background, and then which pixels of the image are objects,
being able to separate close or superimposed objects. For each identified H\,{\sc ii} regions in the NGC~1232 H$\alpha$ image was performed with the aid of  region, 
its position and size are determined by fitting an ellipse function. 
It also derives the photometry of each source.
There are two important input parameters used in SExtractor for 
H\,{\sc ii} region photometry, the minimum number of pixels of a given source and the minimum flux
compared to the background.
We adopted 10 pixels as the minimum number of pixels for
an object to be detected, which roughly correspond to the seeing of the observation. 
We also required that to be detected the sources should have a minimum flux of
2$\sigma$, where $\sigma$ is the average sky noise around the source. 
The average sky noise surface brightness in our image was about 2 $\times$ 10$^{-18}$ erg cm$^2$ s$^{-1}$ arcsec$^{-2}$.

We detected 976 sources. The highest number of sources
ever detected for this galaxy was 529 sources by \citet{hodge+83}.
It is important to take into account that our result almost doubles the number of sources already
detected for this galaxy, even though our image does not encompass the whole galaxy but only the central part.

A direct comparison of individual H\,{\sc ii} regions properties 
from our study and \citet{hodge+83} is extremely difficult.
This is because
their study was done using a photographic plate, and no information about 
sensitivity is given in their paper. Also, they state that  
the individual separation of the objects
is dependent on the properties of the plate, and many times
a number of bright knots in a lower-luminosity envelope were given a single
object identification number. 

\begin{figure*}
\centering
\includegraphics[scale=0.9]{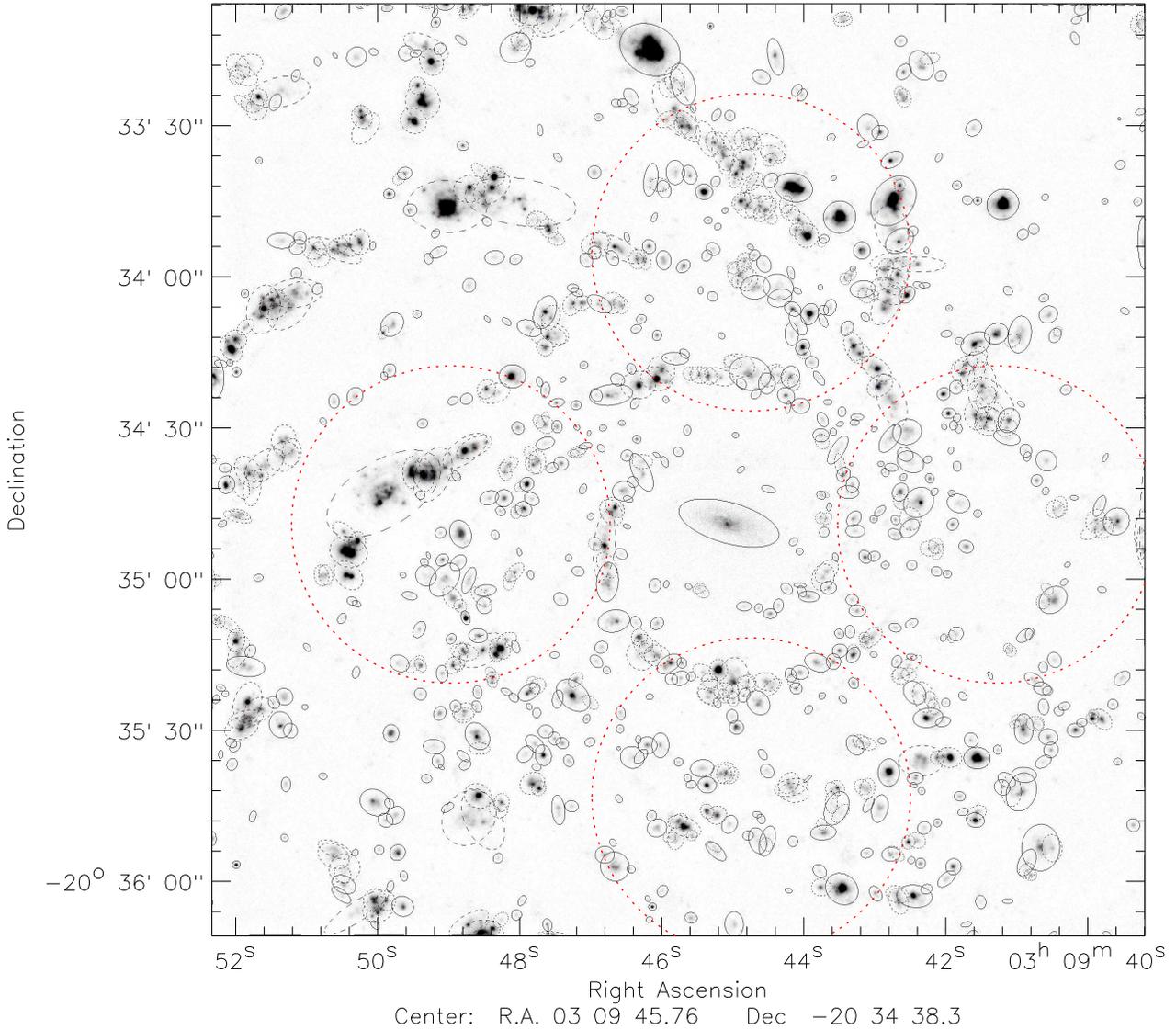} 
\caption{The image shows the H\,{\sc ii} regions fitted by SExtractor for NGC~1232.
As in figure~\ref{ngc1232_filters}, north is up and east is left.
Darker H\,{\sc ii} regions are regions with higher H$\alpha$ emission. 
Black dashed lines represent 
fonts marked with the crowded flag, where the Kron radius measured by SExtractor
was not determined with confidence. We detected 976, which is the highest number of sources ever detected for this galaxy. 
The large dotted red circles refer to figure~\ref{hiilumfunc}}.
\label{ngc1232_sources}
\end{figure*}

In any case, we expect that the higher number of H\,{\sc ii} regions identified here can help to improve studies such as the nature 
of the spiral structure and the spatial distribution of star-forming regions, 
abundance gradients, and the use of H\,{\sc ii} properties (sizes, luminosities, etc.) for the extragalactic
distance scale.
Figure~\ref{ngc1232_sources} shows the sources detected by SExtractor. Table~\ref{sources}
shows, as an example, the ten most luminous H\,{\sc ii} regions detected. The complete catalog 
can be found in the online material. 

\begin{table*}[h]
\caption{Data for the 10 most luminous H\,{\sc ii} regions detected in galaxy NGC~1232. The complete catalog can be found in the on-line material.}
{\renewcommand{\arraystretch}{1.5}
\begin{tabular}{|c|c|c|c|c|c|c|}
\hline

Source & Position (RA) &  Position (Dec) & F(H$\alpha$) (10$^{-17}$erg/$cm^{2}$/s)  & SFR ($M_{\odot}$/yr) & Area (pixel$^2$) & L(H$\alpha$) (erg s$^{-1}$)\\
\hline
 1.0    &  03h09m45.70s    & -20h33m16.35s    & 16220.49 $\pm$     7.99    &      40.332e-3 $\pm$    4.840e-3 &        9670.00    &    7.61e+39\\
\hline
 2.0    &  03h09m42.93s    & -20h33m46.88s    & 11033.00 $\pm$     7.91    &      27.433e-3 $\pm$    3.292e-3 &        9366.00    &    5.18e+39\\
\hline
 3.0    &  03h09m47.64s    & -20h33m43.09s    &  4593.81 $\pm$     4.58    &      11.422e-3 $\pm$    1.371e-3 &        3051.00    &    2.16e+39\\
\hline
 4.0    &  03h09m49.00s    & -20h33m45.82s    &  4528.41 $\pm$     5.67    &      11.260e-3 $\pm$    1.351e-3 &        4841.00    &    2.12e+39\\
\hline
 5.0    &  03h09m42.06s    & -20h34m43.09s    &  3880.59 $\pm$     7.12    &       9.649e-3 $\pm$    1.158e-3 &        7131.00    &    1.82e+39\\
\hline
 6.0    &  03h09m50.50s    & -20h33m46.72s    &  3652.16 $\pm$     4.41    &       9.081e-3 $\pm$    1.090e-3 &        2902.00    &    1.71e+39\\
\hline
 7.0    &  03h09m41.58s    & -20h34m54.59s    &  3640.34 $\pm$     4.34    &       9.052e-3 $\pm$    1.086e-3 &        2848.00    &    1.71e+39\\
\hline
 8.0    &  03h09m48.26s    & -20h33m48.91s    &  2729.06 $\pm$     4.11    &       6.786e-3 $\pm$    0.814e-3 &        2503.00    &    1.28e+39\\
\hline
 9.0    &  03h09m44.12s    & -20h33m08.82s    &  2611.27 $\pm$     3.74    &       6.493e-3 $\pm$    0.779e-3 &        2147.00    &    1.23e+39\\
\hline
 10.    &  03h09m42.61s    & -20h33m26.54s    &  2354.45 $\pm$     4.36    &       5.854e-3 $\pm$    0.703e-3 &        2839.00    &    1.10e+39\\
\hline
\end{tabular} }
\label{sources}
\end{table*}

The H$\alpha$ luminosity of each H\,{\sc ii} region can be obtained from the H$\alpha$ flux
measured by SExtractor, after correcting for two effects: 
internal extinction and the contamination from
[N\,{\sc ii}] lines to the total flux.

For the internal extinction, we adopt the values of \citet{bresolin+05}, who 
measured
emission lines of 13 H\,{\sc ii} regions in NGC~1232.
They obtained the interstellar extinction using the
Balmer decrement measured using the H$\alpha$, H$\beta$, and H$\delta$ lines. From their values of 
c(H$\beta$) and using \citet{cardelli+89} extinction law, we obtain A$_{H\alpha}$,
which varies from 0.0 to 0.8, with an average value of 0.52, removing
the lowest and the highest values. We then applied this average
value to correct the flux of all H\,{\sc ii} regions in our sample.

The
correction for the contamination of [N\,{\sc ii}] is necessary because
the lines [N\,{\sc ii}]~$\lambda$6548\AA~and [N\,{\sc ii}]~$\lambda$6584\AA~are very close to the 
H$\alpha$ line and fall within the bandpass of the H$\alpha$ filter used. To this purpose, we 
again use the values obtained by  \citet{bresolin+05}. 
Although the study is based only on 
strong emission-line regions, searching for high-metallicity objects, we believe they give the best estimate for the
[N\,{\sc ii}]/H$\alpha$ ratio we could get for this galaxy,
instead of doing actual
spectroscopy of each region (which is beyond the scope of this paper). 
They found a large variation of this ratio, with values 
between 0.07 and 0.63, with 61\% of the objects with a ratio
smaller than 0.3. We excluded the highest and lowest values, and
from these data we obtained an average [N\,{\sc ii}]$\lambda6584$\AA/H$\alpha$ 
ratio of 0.24. To account for the other [N\,{\sc ii}] line,
we used the theoretical ratio between the two, which is 1/3. 
We then assumed that, on average, 32\% of the flux measured by SExtractor for each H\,{\sc ii} region
is a contamination by the [N\,{\sc ii}] lines, and corrected for this amount.
It is important to realize that this is an approximation, since 
each object might have very different ratios from this average value,
and the errors associated with this will propagate to the H$\alpha$ luminosity and
SFR measurements.
After these corrections the H$\alpha$ fluxes were employed to derive the
H$\alpha$ luminosity adopting a galaxy distance of 19.8~Mpc.

\section{Catalog Properties}

The faintest H\,{\sc ii} region luminosity measured was about 7.6 $\times$ 10$^{35}$ erg s$^{-1}$. These
low luminosity regions are probably ionized by single stars or 
matter bounded regions with significant photon loss 
\citep{youngblood+99}. The most luminous H\,{\sc ii} region 
we detected in NGC~1232
has a luminosity of 6.6 $\times$ 10$^{39}$ erg s$^{-1}$, comparable with the luminosity of 
the 30 Doradus nebula in the Large Magellanic Cloud \citep[7 $\times$ 10$^{39}$ erg s$^{-1}$,][]{kennicutt84}. 
In this section, we will analyze the general properties of the H\,{\sc ii} regions detected here with respect with
their luminosity and sizes. 

The H\,{\sc ii} luminosity function is an important diagnostic of star formation properties in galaxies
 \citep{kennicutt+89}. 
The differential H\,{\sc ii} luminosity function is  usually parameterized as a power law
\citep{kennicutt+89, banfi+93,oey+98}:

\begin{equation}
N(L)dL = AL^{-a}dL,
\end{equation}

where N(L)dL is the number of nebulae with H$\alpha$ luminosities in the
range of L to L+dL,
{\it A} is a constant and {\it a} is the power-law index. Past studies found that the slope of the H\,{\sc ii} luminosity function is correlated with 
the morphological type, in the sense that early-type galaxies show steeper slopes
than late types. The power-law index {\it a} is $\approx$ 2.6 for Sa galaxies \citep{caldwell+91},
$\approx$ 2.0 for Sb-Sc galaxies \citep{kennicutt+89, banfi+93}, and 
between 1.0 and 1.7 for Im galaxies \citep{kennicutt+89, youngblood+99}.
In a study of the H\,{\sc ii} region luminosity functions of 30
nearby galaxies, \citet{kennicutt+89} found that late-type spirals and
irregulars have shallower slopes often extending beyond 10$^{39}$ ergs s$^{-1}$,
while among earlier-type spirals the luminosity functions rarely extend beyond
this value.
These differences have been interpreted as differences in star formation
properties, like differences in gas dynamics and molecular cloud mass spectrum 
\citep{ kennicutt+89, thronson+91, rand92} 
or related to the evolution of the ionizing clusters, where steeper slopes would
be the consequence of aging effects \citep{oey+98}. 
Steeper slopes 
in earlier-type galaxies can be explained by a lower maximum number
of ionizing stars cutoff found for the parent OB associations
\citep{oey+98}. 
Late-type galaxies would have a large number of what 
\citet{oey+98} calls ''saturated" clusters, which are rich clusters with
complete statistics in terms of initial mass function. These clusters
would be responsible for the high-luminosity end of the luminosity function.

\begin{figure*}
\centering
\includegraphics[width=16.cm]{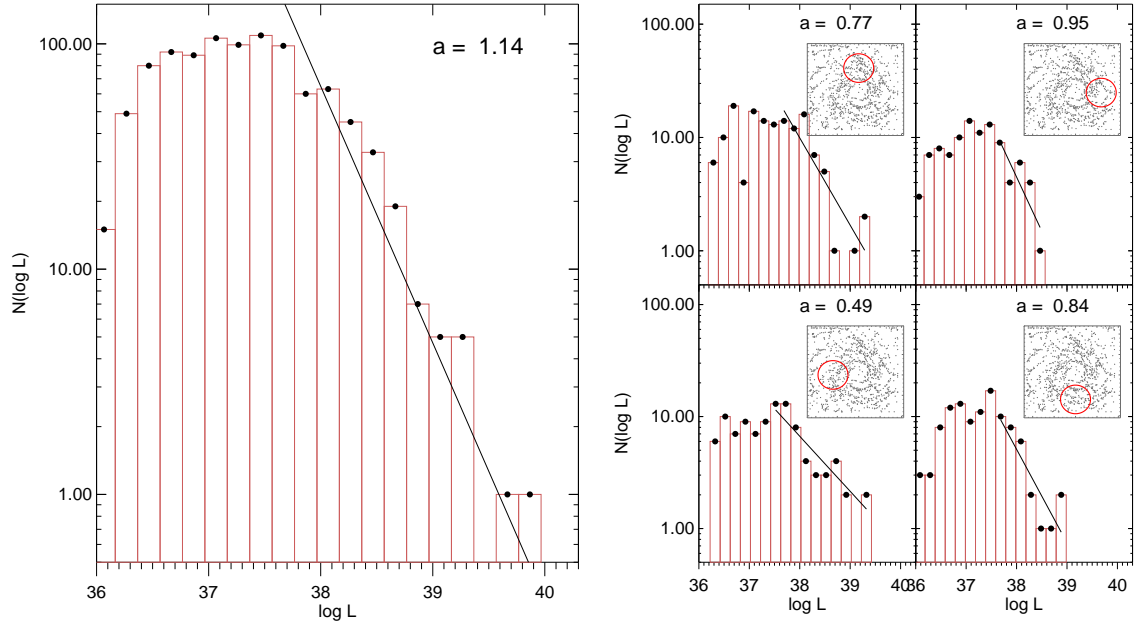} 
\caption{H\,{\sc ii} luminosity function of NGC~1232. The left panel shows the luminosity function including
all H\,{\sc ii} regions of the galaxy. The solid line represents the power-law fitted for log L $\geq$ 37.5 dex. 
The four panels in the left side of this figure correspond to the luminosity functions 
in four different regions of the galaxy, as indicated by the red circle
over the galaxy's image of each plot. Slope values a for each of the fits are shown in the top of each plot. }
\label{hiilumfunc}
\end{figure*}

We built the luminosity function for NGC~1232, following the same
criteria used by previous authors
 \citep[e.g.][]{kennicutt+89,oey+98,youngblood+99}, 
binning the luminosities in logarithmic intervals
of 0.2 dex. The result is shown in figure~\ref{hiilumfunc}. 
It can be seen that the luminosity function in NGC~1232 shows a break 
in slope, with the fainter H\,{\sc ii} regions showing a shallower slope
compared with the high-luminosity objects.
This same effect was already observed for a large 
number of other galaxies
\citep[e.g.][]{kennicutt+89,rand92,walterbos+92,rozas+96}. \citet{oey+98} showed
that this might be related to the aging of the stellar population, where
the slope transition moves to lower L as the population ages and objects grow fainter. 
\citet{kennicutt+89} found that for most galaxies in their sample, this turnover
point occurred between 10$^{36}$ and 10$^{37}$ ergs s$^{-1}$. 
In NGC~1232 the break in slope happens around log L $=$ 37.5 dex. To obtain the 
power law index, we excluded the points below this value. With that we obtained 
a $=$ 1.14 for the power law index (left panel of  figure~\ref{hiilumfunc}). This
value is lower than typical values found for Sc galaxies, being closer to the values found 
for Im galaxies. In agreement with what is seen in the literature, where shallower 
luminosity functions are related with the presence of high-luminosity H\,{\sc ii} regions, we
find that NGC~1232 has a significant 
number of H\,{\sc ii} regions with luminosities
higher than 10$^{39}$ ergs s$^{-1}$. 

The properties of the H\,{\sc ii} regions could be affected by dynamical and chemical differences
along the disk 
\citep[e.g.][]{hodge87,knapen+98,cedres+13}.
For this reason, it is important to determine if the luminosity function is 
affected by the location of the H\,{\sc ii} regions in the galaxy.
This is especially important in the
scenario where a possible collision with a dwarf galaxy might have affected different regions
of this galaxy in different ways. 
We then constructed the luminosity function in four different parts of the galaxy, all at the
same distance from the center
($\approx$ 2.7 kpc), one in each direction from the nucleus, in a way that the west region of the 
galaxy that includes the x-ray peak was inside one of them. 
This is shown in the right 
panel of figure~\ref{hiilumfunc}, where for each plot the red circle over the image of the galaxy represents
the region considered.  
From this plot we can see that the western region of the galaxy has no high-luminosity H\,{\sc ii} regions,
which makes the slope of the luminosity function steeper, compared to the other regions.
The high-luminosity regions are mostly concentrated in the northern and in the eastern regions
of the galaxy. 

\begin{figure}
\centering
\includegraphics[width=12.cm]{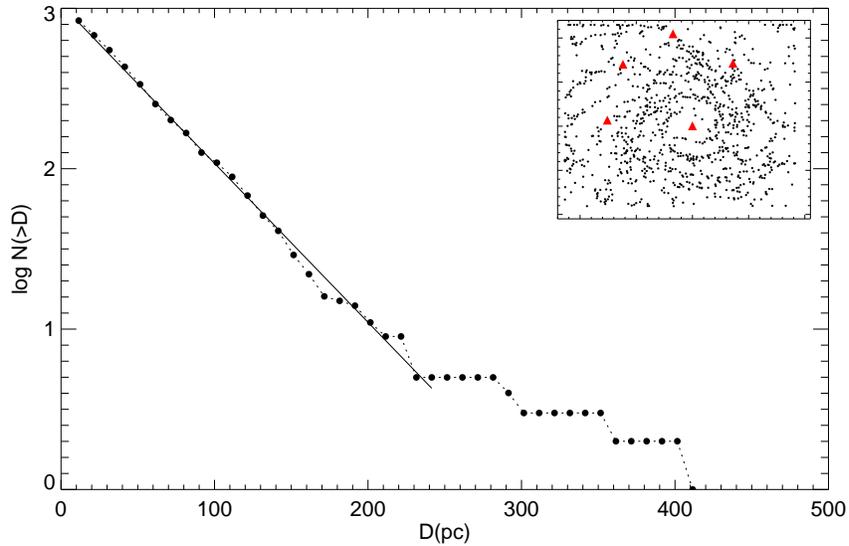} 
\caption{H\,{\sc ii} regions 
size cumulative distribution. The inset in the top right
of this figure shows the location of the five brightest H\,{\sc ii} regions detected
(red triangles), which
were excluded from the the fit. The fit to the remaining regions is shown by the solid line. 
The characteristic diameter of the galaxy obtained by this fit is  D$_0$ $=$ 100.9$\pm$ 1.5 pc.}
\label{hiidiadist}
\end{figure}

With respect to sizes, \citet{vandenbergh81} proposed that the frequency distribution of H\,{\sc ii} regions in
spiral galaxies follows the law

\begin{equation}
N = N_0 e^{-D/D_{0}},
\end{equation}

where N is the number of H\,{\sc ii} regions observed with a diameter larger than D, 
and D$_{0}$ is the characteristic diameter of the galaxy. 
N$_{0}$ is an (extrapolated) characteristic value for the total number of regions.
Most spiral galaxies, independent of the specific morphological type, seem to follow
this distribution \citep{Ye92}.
We obtained the 
diameters of the H\,{\sc ii} regions of NGC~1232 from the areas measured by SExtractor, 
defining D = $2(A/\pi)^{1/2}$, where A is the area, and constructed this cumulative
distribution function, as shown in figure~\ref{hiidiadist}. From this figure
it is possible to see that the largest regions are too big compared to the 
distribution defined by the rest of the regions. This effect was previously
observed for other galaxies as well \citep{hodge+83, youngblood+99}. 
These points, all located in the upper part of the galaxy, were excluded when determining the slope of the diameter
distribution. We then obtained D$_0$ $=$ 100.9$\pm$ 1.5 pc for 
NGC~1232,
which is a typical value found for  spiral galaxies \citep{hodge87,Ye92}.

\begin{figure}
\centering
\includegraphics[width=12.cm]{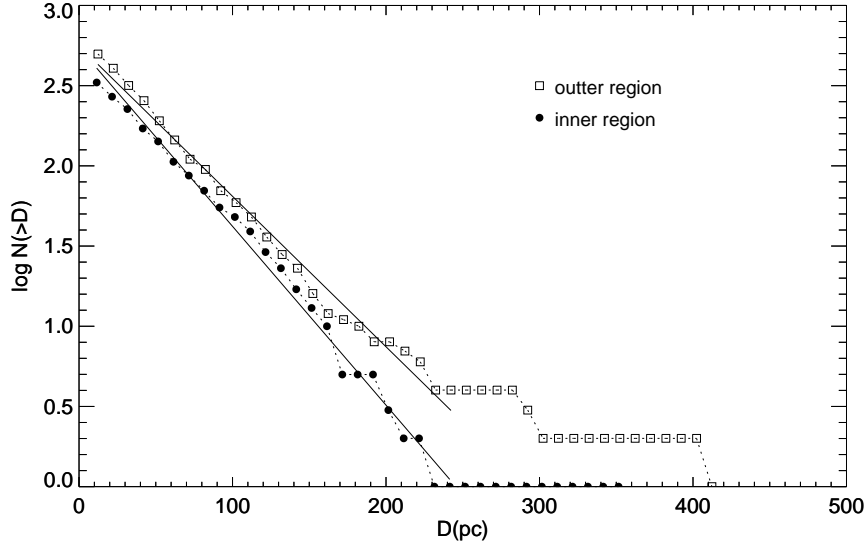} 
\caption{H\,{\sc ii} regions 
size cumulative distribution for the inner (filled circle) and outer (open square)
region of NGC~1232. 
The inner region is defined as the region inside half of the maximum H\,{\sc ii} region
distance to the center, which is 11.2 kpc. 
The solid lines represent the fits for each distribution.
We found D$_0$ $=$ 89.7 $\pm$ 1.7 pc for the inner region and D$_0$ $=$ 106.6 $\pm$ 2.6 pc
for the outer region, following the trend found in the literature, which shows
a tendency to have a larger D$_0$ for outer than for inner regions.}
\label{hiidiadep}
\end{figure}

\citet{hodge87} suggested that there is a variation of the diameter distribution with 
galactocentric distance, showing a tendency to have a larger D$_0$ for outer than for
inner regions. 
 Their results imply that the slope of the diameter distribution might be related to
the environment, probably the gas density and/or the dynamical environment of the gas.
In order to verify this for NGC~1232, we divided it in two parts: inner and outer region.
The inner region is defined as the region inside half of the maximum H\,{\sc ii} region distance to the center, which is
11.2 kpc. Indeed, we found D$_0$ $=$ 89.7 $\pm$ 1.7 pc for the inner region and D$_0$ $=$ 106.6 $\pm$ 2.6 pc
for the outer region. The different cumulative distributions for the inner and outer regions
are shown in figure~\ref{hiidiadep}. From this figure we can also see that the five largest
H\,{\sc ii} regions are located in the outer region of NGC~1232. 

\begin{figure}
\centering
\includegraphics[width=12.cm]{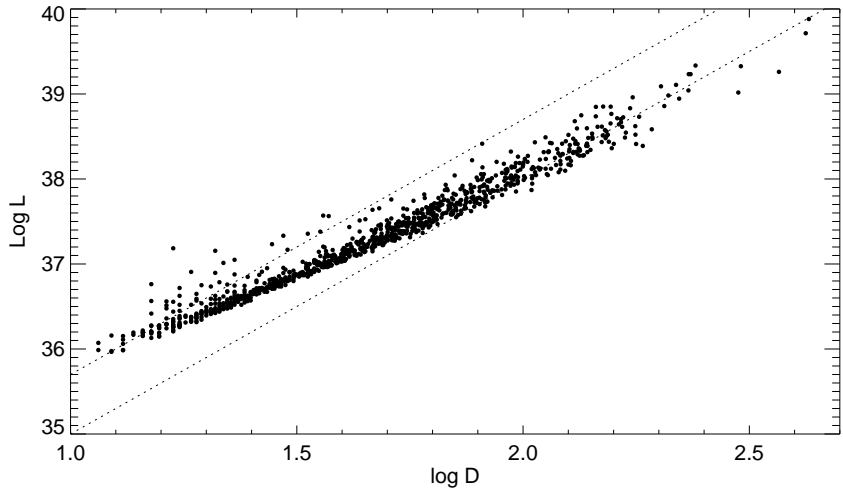} 
\caption{Logarithm of the Luminosity versus the logarithm of the diameter for the H\,{\sc ii} regions
detected in NGC~1232. The dotted lines indicate the trend 
expected if the H\,{\sc ii} regions are radiation
bounded,
which is the cube root of the ionizing luminosity}. 
\label{hiidialum}
\end{figure}

Another characteristic of the H\,{\sc ii} regions that can be explored is if they are radiation
or matter bounded. If they are radiation bounded and have similar gas densities across the
galaxy, then their diameter should scale as the cube root of the ionizing luminosity
\citep{osterbrock+89}. 
Previous studies \citep[e.g.][]{kennicutt+89, banfi+93} found that H\,{\sc ii} regions in galaxies
are essentially radiation bounded. To verify that for the  H\,{\sc ii} regions of NGC~1232 
we plotted the luminosities versus their diameters. This is shown in figure~\ref{hiidialum}.
In this figure the two dotted lines mark the 
expected slopes for radiation bounded
H\,{\sc ii} regions. 
It is important to realize that most of these regions will be ionized by
more than one star, which results in a situation physically different from that 
expected by a Str\"ongren sphere. However, we can see from figure~\ref{hiidialum}
that, at least the larger regions seem to be indeed radiation bounded. Smaller regions,
as previously mentioned, might be matter bounded.

\section{Star Formation Rate}

The SFR is then obtained adopting the equation from \citet{calzetti+07}:

\begin{equation}
SFR(M_\odot~yr^{-1}) = 5.3 \times 10^{-42} L(H\alpha)~~(erg~ s^{-1}).
\end{equation}

\begin{figure*}
  \centering
    \includegraphics[width=8.3cm]{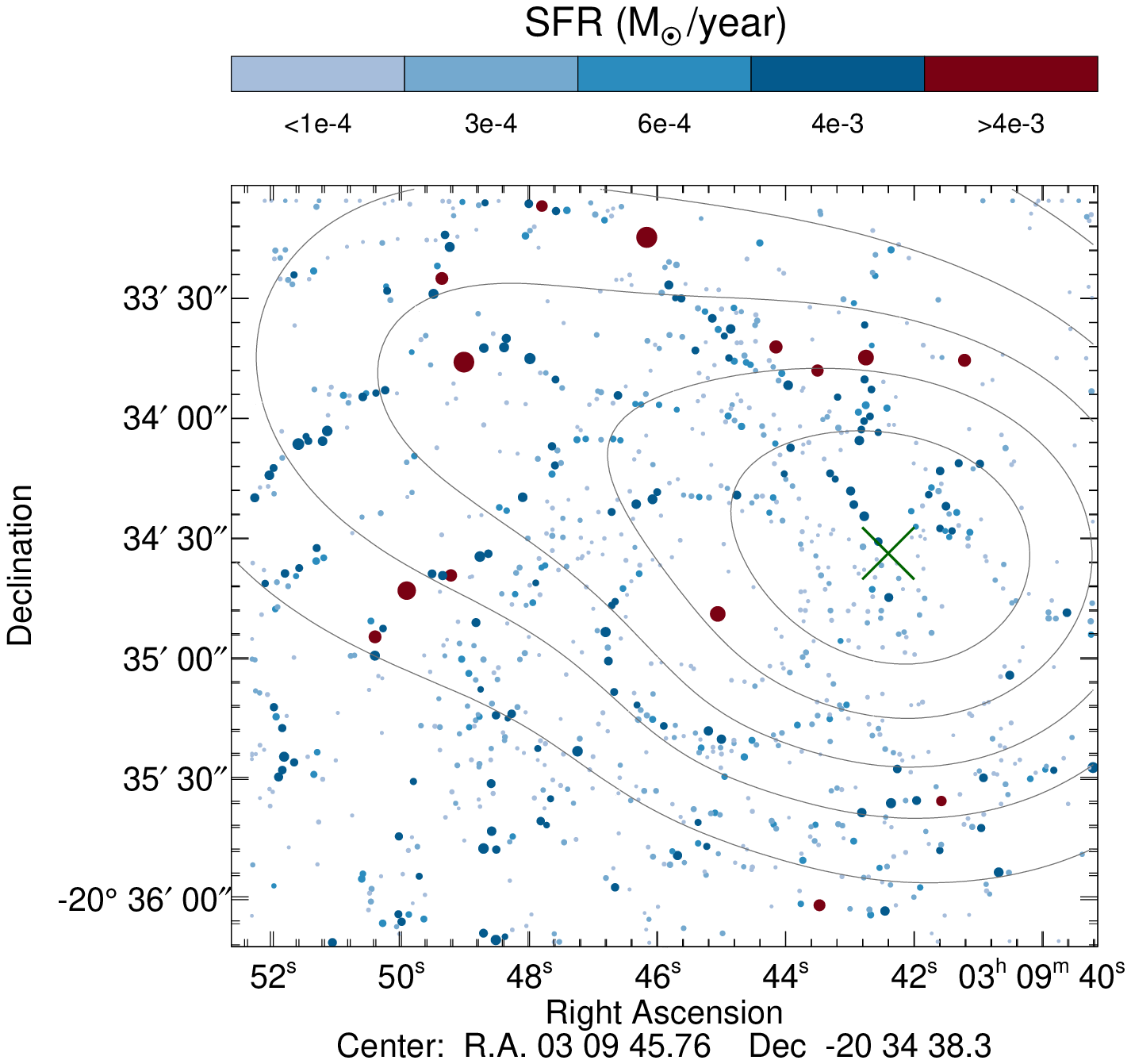} 
    \qquad
   \includegraphics[width=8.3cm]{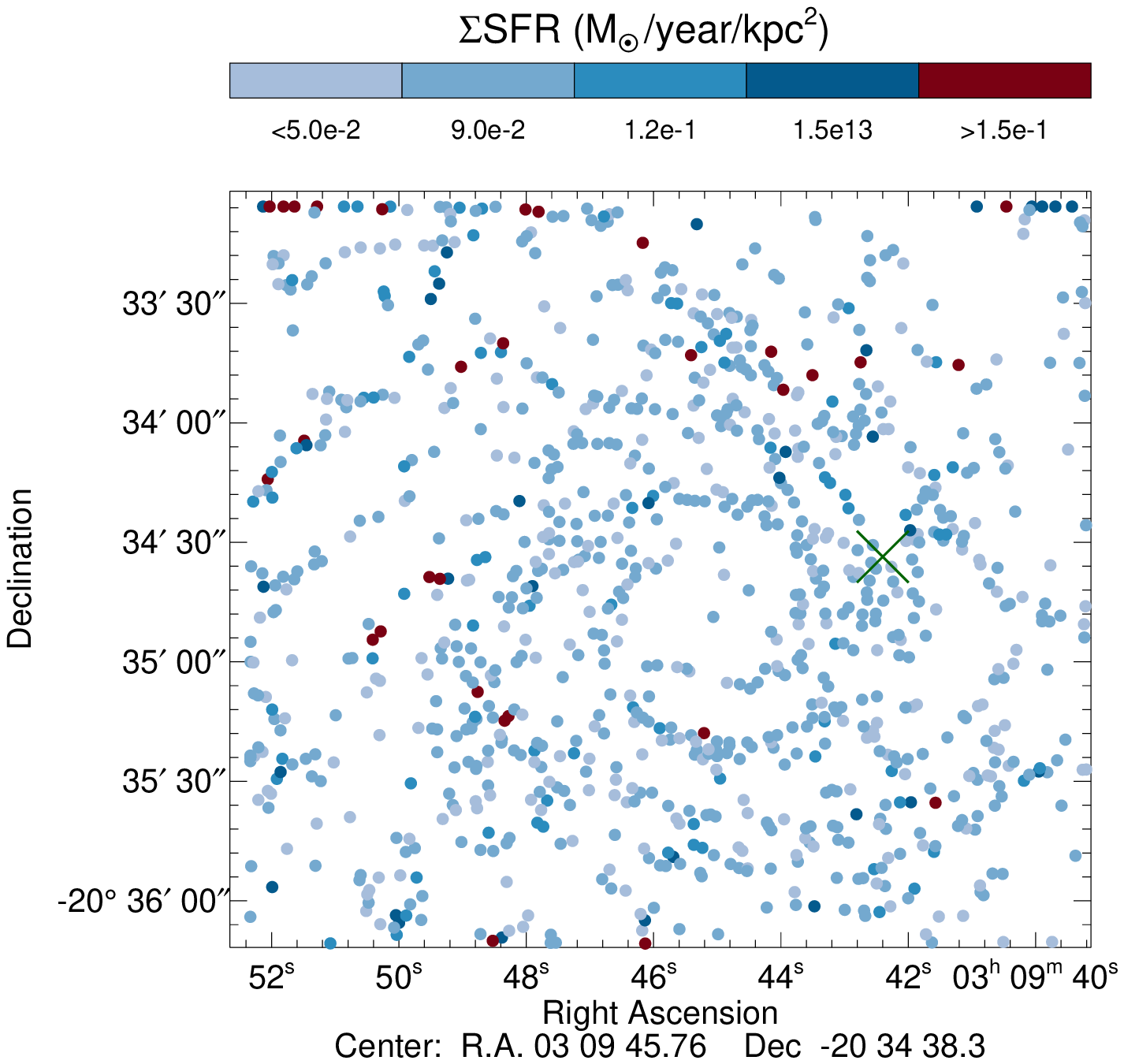} 
    \caption{The SFR of NGC~1232, calculated for each H\,{\sc ii} region. Left: SFR distribution. The 
size of the dots is proportional to the size of each H\,{\sc ii} region. 
The gray contour represents the x-ray contours from \citet{garmire+13}.
Right: distribution of the SFR 
per unity area. The green  ``x" in both plots marks the peak of the x-ray emission 
(X-ray data obtained from G. Garmire, private communication.)}
    \label{sfrdistrib}%
\end{figure*}

The H$\alpha$ flux, luminosity, and the SFR for each H\,{\sc ii} region are
presented in table~\ref{sources}. The left side of figure~\ref{sfrdistrib} shows the SFR distribution
in the galaxy. 
The uncertainties in the SFR vary from 12\% to 17\%, with an average value of 13\%, taking into
account the errors in the fluxes obtained by SExtractor and the error in the
distance determination. It does not take into account the approximations for the extinction and [N\,{\sc ii}]
correction, which are much harder to estimate.

As expected, the SFR 
is higher in the spiral arms, as is commonly accepted in the 
literature that the H\,{\sc ii} regions are tracers of these arms. Another quantity that can
be studied with our data and which carries more physical meaning, is the SFR density, or SFR per
unity area ($\Sigma$SFR). The results are shown in the right panel of figure~\ref{sfrdistrib}. 
$\Sigma$SFR was obtained by dividing the SFR of every H\,{\sc ii} region by their physical size,
given by SExtractor. In can be seen in the right panel of figure~\ref{sfrdistrib} that the spiral pattern becomes 
less obvious, 
supporting the hypothesis that the spiral arms only show more star formation because they concentrate most of the gas, 
but not necessarily are they more efficient at forming stars 
\citep{scoville+01, foyle+10,gutierrez+11,kreckel+16}.
Of course, a more conclusive result can only be achieved by separating the H\,{\sc ii} regions of the
arms from the inter-arms, which is 
 non-trivial for multi-armed galaxies.
 
In terms of general behavior, NGC~1232 is a typical spiral galaxy, which have total SFR values
around 1 $M_\odot~yr^{-1}$ \citep{lee+09}. 
We found a value of 
0.77 $M_\odot~yr^{-1}$ from our data, measuring the total H$\alpha$ flux inside
a large aperture and performing the same process to obtain the H$\alpha$ luminosity as was done for each 
H\,{\sc ii} region. 
It is important to remember that our data do not encompass the whole galaxy. 
Flux sensitivity also has to be taken into account for direct comparisons with
other authors, in the sense that low sensitivities will result in significant amounts
of H$\alpha$ diffuse emission not being detected, leading to a lower SFR measurement.
Our sensitivity is just slightly higher than that of \citet{lee+09}.

The $\Sigma$SFR values obtained in this work for NGC~1232 can be compared with the values
obtained by \citet{grosbol+12}. They derived $\Sigma$SFR gradients for several spiral galaxies,
including NGC~1232, finding an essentially constant $\Sigma$SFR along the radius of the galaxy,
with values between 10$^{-2}$ - 10$^{-3}$ $M_\odot~yr^{-1} kpc^{-2}$. With our superior image quality, 
we confirm their results. 
If we look at the average radial values, we can also
see that there is no obvious $\Sigma$SFR gradient in the galaxy.

However, the most interesting feature that we found in this distribution is a clear 
asymmetry in the intensity of the SFR, where the regions of more intense formation (red symbols) 
appear to be located preferably in the northeast side of the galaxy
(figure~\ref{sfrdistrib}), while we notice an absence
of star formation to the west and very likely to the south of the nucleus. 

As mentioned previously, \citet{garmire+13} found a diffuse X-ray emission in NGC\,1232
that indicates a possible collision with a dwarf galaxy.
The peak of the diffuse 
X-ray emission is coincident with the zone where the H\,{\sc ii} regions are 
smaller and more sparse, while
the excess SFR in the northeast of the galaxy is where the emission is weaker. 
The X-ray contours in the left panel of figure~\ref{sfrdistrib} can help to see that. 
We interpret this result as a suppression of the star formation in the hot gas region and/or as
an intensification of the star formation in the post-shock region of the dwarf galaxy's passage.

\begin{figure}
\centering
\includegraphics[width=13.cm]{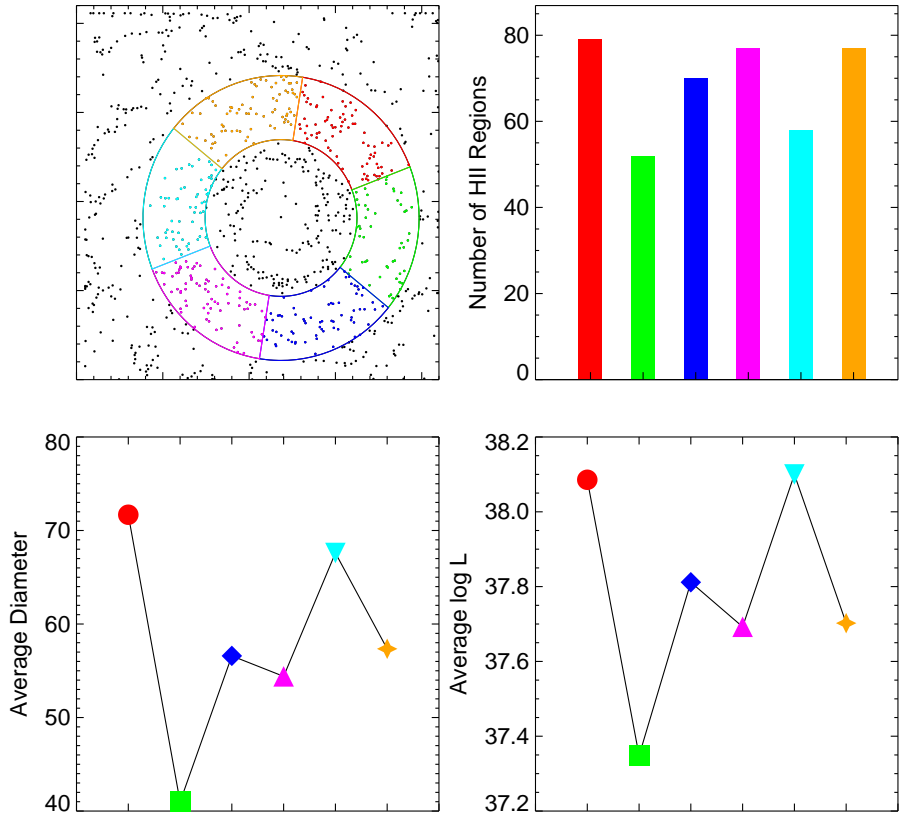} 
\caption{Main properties of the H\,{\sc ii} regions in six different regions of NGC~1232, at the same distance from the center, 
color-coded as shown in the top left plot. The top right plot shows the number of H\,{\sc ii} regions in each section,
starting by the one on the top left (red region) and with following in the clockwise direction. 
The bottom left plot shows the average diameter of the H\,{\sc ii} regions (in pc) in each section as measured by SEXTRACTOR 
and the bottom right plot the average luminosity (in Log units), following the same order as the top right plot. It can be
clearly seen that the green region (second point in each plot), which is around the region of the X-Ray peak, has a deficiency of 
large and luminous H\,{\sc ii} regions.
}
\label{corrfunct}
\end{figure}

To make this effect more evident, 
we picked an annulus including the x-ray peak region, and divided 
this annulus in six sections of 60$^o$ each. For each of these sections we computed average 
properties related to the H\,{\sc ii} regions, namely, the number of regions in each section, 
the average diameter of the regions, and average luminosity. This is shown in 
figure~\ref{corrfunct}. These numbers will, of course, be affected by the size of the chosen regions (which, if
too big will just average out any localized differences) and their placement
(inter-arm locations will naturally have less H\,{\sc ii} regions than in arm locations).
By looking at sections at the same distance from the center in all directions we tried to avoid both 
these effects. It can be seen from Figure~\ref{corrfunct} that the green section is clearly
deficient compared to the other regions, both in terms of number of regions and in
average properties (sizes and luminosities). This section is located close to the 
x-ray peak. 
This result provides additional evidence on the quenching of the star formation as a 
result of the collision with the dwarf galaxy. In-depth spectroscopic study will certainly 
be helpful to reach a definitive conclusion on this issue.

\section{Conclusions}

In this work, we study the star formation distribution in the galaxy NGC~1232, a spiral galaxy seen 
practically face on. It is considered a grand-design spiral, 
and therefore an excellent laboratory for the study of the relationship between star formation and 
spiral arms. Besides that, \citet{garmire+13} reports evidence that a dwarf galaxy might have
crossed the disk of NGC~1232, based on the detection of a diffuse, hot X-ray emission cloud
observed in this galaxy. 

We obtained a high spatial resolution H$\alpha$ image of NGC~1232 using SAM at SOAR telescope, with AO. This resulted in the best spatial resolution image of this galaxy so far. 
A total of 976 H\,{\sc ii} regions were detected for the SOAR image field, which encompasses about the 50\%
central part of optical image of the galaxy. Despite not covering the entire galaxy,  
the number of sources detected here represent the most complete H$\alpha$ source catalog 
for this galaxy in the literature, since double the number of H\,{\sc ii} regions were already
detected in this object. 

We also constructed the H\,{\sc ii} regions luminosity function, obtaining a power-law index
of a = 1.14. This is a value typically lower than what is found for Sc galaxies. In the case
of NGC~1232 this is related to the presence of a significant number of high-luminosity
H\,{\sc ii} regions. These are mainly found in the northern and eastern part of the galaxy,
while the eastern part of the galaxy clearly shows a lack of these brighter regions. We also
constructed the size distribution function of the H\,{\sc ii} regions, verifying that,
as for most galaxies, NGC~1232 follows an exponential law, with a characteristic diameter
of D$_0$$=$100.9 $\pm$ 1.5 pc. 

We used the H$\alpha$ luminosity to determine the SFR of each H\,{\sc ii} region, and analyzed their
distribution in the galaxy. We also calculated the SFR density $\Sigma$SFR for each 
source. Results show that, as expected, stronger star formation is found along the spiral
arms. However, the apparent concentration of H\,{\sc ii} regions in the spiral arms is diluted 
when the SFR density is analyzed. 

We also found an interesting pattern in the distribution of the SFR, where there seems to be 
an lower number of H\,{\sc ii} sources exactly in the region where the diffuse X-ray gas 
has a peak of emission.
This might be due to a quenching of star formation due to the collision with the dwarf galaxy. On the other
hand, there seems to be an excess of star formation in the northeast part of the galaxy,
where the X-ray emission is weaker. We suggest that this excess might have been induced by a 
shock wave from the collision.

\acknowledgments
We thank the referee for invaluable suggestions that greatly improved the paper. 
Based on observations obtained at the Southern Astrophysical Research (SOAR) telescope, which is a joint project of the Minist\'{e}rio da Ci\^{e}ncia, Tecnologia, e Inova\c{c}\~{a}o (MCTI) da Rep\'{u}blica Federativa do Brasil, the U.S. National Optical Astronomy Observatory (NOAO), the University of North Carolina at Chapel Hill (UNC), and Michigan State University (MSU).
A.A.S. acknowledges CAPES for financial support.
L.M. thanks CNPQ for financial support through grant 303697/2015-6 and FAPESP through
grant 2015/14575-0. 
A.R.A. thanks CNPq for partial
support to this work.
We thank Andrei Tokovinin from the SOAR observatory, for the excellent support during the observations with SAM/SOAR.
We also thank G. Garmire for the x-ray data used in this paper.

\vspace{5mm}
\facilities{SOAR(SAM)}

\software{IRAF \citep{valdes98,valdes+98},
          SExtractor \citep{bertin+96}
          }

\appendix
\section{Sources Data}

\startlongtable
\begin{deluxetable*}{ccccccc}
\tablecaption{Data from sources detected in galaxy NGC~1232, in order of luminosity.}
\tablehead{
\colhead{Source} & \colhead{Position (RA)} &\colhead{Position (Dec)} & 
\colhead{F(H$\alpha$) (10$^{-17}$erg/$cm^{2}$/s)}& \colhead{SFR ($M_{\odot}/yr$} & 
\colhead{Area (pixel$^2$)} & \colhead{L(H$\alpha$) (erg s$^{-1}$)}
}
\startdata
\hline
 1.0    &  03h09m45.70s    & -20h33m16.35s    & 16220.49 $\pm$     7.99    &      40.332e-3 $\pm$    4.840e-3 &        9670.00    &    7.61e+39\\
\hline
 2.0    &  03h09m42.93s    & -20h33m46.88s    & 11033.00 $\pm$     7.91    &      27.433e-3 $\pm$    3.292e-3 &        9366.00    &    5.18e+39\\
\hline
 3.0    &  03h09m47.64s    & -20h33m43.09s    &  4593.81 $\pm$     4.58    &      11.422e-3 $\pm$    1.371e-3 &        3051.00    &    2.16e+39\\
\hline
 4.0    &  03h09m49.00s    & -20h33m45.82s    &  4528.41 $\pm$     5.67    &      11.260e-3 $\pm$    1.351e-3 &        4841.00    &    2.12e+39\\
\hline
 5.0    &  03h09m42.06s    & -20h34m43.09s    &  3880.59 $\pm$     7.12    &       9.649e-3 $\pm$    1.158e-3 &        7131.00    &    1.82e+39\\
\hline
 6.0    &  03h09m50.50s    & -20h33m46.72s    &  3652.16 $\pm$     4.41    &       9.081e-3 $\pm$    1.090e-3 &        2902.00    &    1.71e+39\\
\hline
 7.0    &  03h09m41.58s    & -20h34m54.59s    &  3640.34 $\pm$     4.34    &       9.052e-3 $\pm$    1.086e-3 &        2848.00    &    1.71e+39\\
\hline
 8.0    &  03h09m48.26s    & -20h33m48.91s    &  2729.06 $\pm$     4.11    &       6.786e-3 $\pm$    0.814e-3 &        2503.00    &    1.28e+39\\
\enddata
\end{deluxetable*}

\bibliographystyle{aasjournal}
\bibliography{bib}

\end{document}